\documentclass[runningheads,a4paper]{llncs}
\usepackage{color}
\usepackage{amssymb}
\usepackage{graphicx}
\usepackage{framed}
\usepackage{url}
\usepackage{float}
\usepackage{caption}
\usepackage{subfig}
\urldef{\mailsa}\path|firstname.lastname@gesis.org|
\newcommand{\keywords}[1]{\par\addvspace\baselineskip
\noindent\keywordname\enspace\ignorespaces#1}
\begin{document}

\mainmatter  % start of an individual contribution

% first the title is needed
\title{Analyzing the network structure and gender differences among the members of the Networked Knowledge Organization Systems (NKOS) community}

% a short form should be given in case it is too long for the running head

% the name(s) of the author(s) follow(s) next
%
% NB: Chinese authors should write their first names(s) in front of
% their surnames. This ensures that the names appear correctly in
% the running heads and the author index.
%
\author{Fariba Karimi, Philipp Mayr \and Fakhri Momeni}
\titlerunning{Network structure of the NKOS community}
% (feature abused for this document to repeat the title also on left hand pages)

% the affiliations are given next; don't give your e-mail address
% unless you accept that it will be published
\author{Fariba Karimi, Philipp Mayr and Fakhri Momeni}
\institute{GESIS -- Leibniz Institute for the Social Sciences,\\
	Unter Sachsenhausen 6-8\\
	50667 Cologne, Germany\\
	\email{firstname.lastname@gesis.org} }

%
% NB: a more complex sample for affiliations and the mapping to the
% corresponding authors can be found in the file "llncs.dem"
% (search for the string "\mainmatter" where a contribution starts).
% "llncs.dem" accompanies the document class "llncs.cls".
%

%\toctitle{Lecture Notes in Computer Science}
%\tocauthor{Authors' Instructions}
\maketitle

%Deadline: 14-03-20183

\begin{abstract}
In this paper, we analyze a major part of the research output of the Networked Knowledge Organization Systems (NKOS) community in the period 2000 to 2016 from a network analytical perspective. We focus on the papers presented at the European and U.S. NKOS workshops and in addition four special issues on NKOS in the last 16 years. For this purpose, we have generated an open dataset, the "NKOS bibliography" which covers the bibliographic information of the research output. We analyze the co-authorship network of this community which results in 123 papers with a sum of 256 distinct authors.
We use standard network analytic measures such as degree, betweenness and closeness centrality to describe the co-authorship network of the NKOS dataset. 
First, we investigate global properties of the network over time. Second, we analyze the centrality of the authors in the NKOS network. Lastly, we investigate gender differences in collaboration behavior in this community. Our results show that apart from differences in centrality measures of the scholars, they have higher tendency to collaborate with those in the same institution or the same geographic proximity. We also find that homophily is higher among women in this community. Apart from small differences in closeness and clustering among men and women, we do not find any significant dissimilarities with respect to other centralities.  

%The NKOS network is a typical co-authorship network with one large connected component, some smaller components and many isolated co-authorships or triples. The complete NKOS network consists of 97 (38\%) women and 157 (62\%) number of men and 2 unidentified names. The largest component represents 107 authors (41\% of all authors). In the largest connected component we find 46 women and 59 men. \fariba{not suited for the abstract}

\end{abstract}

\section{Introduction}\label{intro}

The Networked Knowledge Organization Systems (NKOS)\footnote{For an introduction of KOS and NKOS and recent applications see \cite{Hill2001,Mayr2016}.} community in Europe and in the United States of America has held a long-running series of annual workshops at the European Conference on Digital Libraries (ECDL), latterly renamed as the International Conference on Theory and Practice of Digital Libraries (TPDL), the Joint Conference on Digital Libraries (JCDL) and some other scattered events. The NKOS workshops in the U.S. have started in 1997/1998 organized by Linda Hill, Gail Hodge, Ron Davies and others. Slightly later, the first NKOS workshop was organized in Europe at ECDL 2000 in Lisbon (Portugal) by Martin Doerr, Traugott Koch, Douglas Tudhope and Repke de Vries.

Typically, recent advances in Knowledge Organization Systems (KOS) have been reported at the annual NKOS workshops, e.g. including the Simple Knowledge Organization System (SKOS) W3C standard, the ISO 25964 thesauri standard, the CIDOC Conceptual Reference Model (CRM), Linked Data applications, KOS-based recommender systems, KOS mapping techniques, KOS registries and metadata, social tagging, user-centered issues, and many other topics\footnote{Comprehensive review articles on KOS and NKOS topics have been published in \cite{Zeng2004,hodge2000}.}. Special issues on Networked Knowledge Organization Systems have been published in Journal of Digital Information in 2001 \cite{Hill2001} and 2004 \cite{Tudhope2004}, in New Review of Hypermedia and Multimedia in 2006 \cite{Tudhope2006} and recently in the International Journal of Digital Libraries in 2016 \cite{Mayr2016}.  Recently, the NKOS workshop activities have accelerated again e.g. with two European NKOS in 2016 at the TPDL and Dublin Core conference and a revival of the U.S. NKOS activities in 2017. In addition, the last two NKOS workshops at TPDL have resulted in formal conference proceedings published as CEUR Workshop Proceedings \cite{NKOS_proceedings_2016,NKOS_proceedings_2017}.

The motivation of this paper is to analyze and visualize the collaboration network of the NKOS community. We are focusing here on the informal part of this output, the paper presentations given at the past NKOS workshops. The specialty of this research output is that these research papers typically are not published in journals or conference proceedings. These papers appear just as oral presentations at the workshop and are documented on the corresponding websites. To cover this informal research output, we have collected presentation information from the workshop agendas. To analyze the co-authorship network of this community, we restrict our analysis to papers which have been authored by a minimum of two authors. This results in 123 papers with a sum of 256 distinct authors. It is important to note that practices at the NKOS workshops in the United States and Europe are different. In the United States, NKOS workshops were previously not based on an open call for papers contribution type, but rather via inviting speakers. This practice explains the relatively low ratio of co-authorship in the U.S. workshop series. From the beginning, in Europe, the NKOS workshops were based on accepting academic papers and resulted in an open call for papers and subsequent peer review of submitted paper abstracts. 

In the following, we report about the network structure and gender differences among the members of the NKOS community as we could recall from the past European and U.S. workshop agendas and published special issues.

This paper is largely an extended version of the paper "Analyzing the research output presented at European Networked Knowledge Organization Systems workshops (2000-2015)" \cite{momeni_2016} presented at the 15th NKOS workshop at TPDL 2016. In \cite{momeni_2016}, we focused on the European workshops and special issues. Meanwhile, we have extended the dataset and included the U.S. NKOS workshops and some other scattered NKOS events. So, this paper is able to give a more comprehensive overview of the international NKOS research community. To the best of our knowledge, this paper is the first attempt to analyze the co-authorship network of NKOS in great details.

In the following sections we describe the underlying dataset (section \ref{dataset}), we perform network analysis (section \ref{analysis}), highlight some results of our analysis (section \ref{results}) and conclude our paper (section \ref{concl}).

\section{NKOS workshop bibliography dataset}\label{dataset}

For our analysis, we have compiled an open dataset derived from the "NKOS bibliography"\footnote{The NKOS workshop bibliography is maintained in the following github repository: https://github.com/PhilippMayr/NKOS-bibliography.}. The NKOS bibliography has been started in 2016 \cite{momeni_2016} and covers bibliographic information of all research papers presented at the past NKOS workshops. Editing, organizing activities (incl. the introductions) at the workshops have not been covered in our dataset. Journal papers published in four special issues on NKOS which have been edited by members of the NKOS community in the same period have been added. These journal papers are the only formal publications in our analysis. In the end, we manually disambiguate author names of all papers. The bibliography is stored in single bibtex files (one bibtex file for each venue).

To this date, the NKOS bibliography covers:
\begin{itemize}
\item sixteen European NKOS workshops from 2000 to 2016. In total 16 workshop agendas: ECDL 2000, 2003-2010, TPDL 2011-2016, Dublin Core 2016,
\item eight U.S. NKOS workshop agendas: JCDL 2000-2003, 2005 and NKOS-CENDI 2008-2009, 2012,
\item four special issues on NKOS \cite{Hill2001,Tudhope2004,Tudhope2006,Mayr2016} and
\item two scattered NKOS workshops at ISKO-UK 2011 and ICADL 2015.

\end{itemize}

For the analysis in this paper, we have compiled all research presentations at NKOS workshops and papers published in special issues. We restrict our analysis to papers which have been authored by a minimum of two authors. This restriction reduces the content of the dataset, e.g. the ECDL NKOS workshop from 2000 is missing in Table \ref{tab:workshops} because all papers were single author papers. In total, this results in a dataset of 123 papers with a sum of 256 
distinct authors (see Table \ref{tab:workshops})\footnote{The data  for this subset is available under https://github.com/PhilippMayr/NKOS-bibliography/tree/master/publications/ijdl17}. 

\begin{table}
	\centering
	\begin{tabular}	{|c|c|c|c|c|}		
		\hline 
        year & nr. papers & nr. authors & nr. links & avg. clustering \\
        \hline
		2001 & 4 & 9 & 6 & 0.37 \\
        \hline
		2002 & 3 & 10 & 13 & 0.8 \\
        \hline
		2003 & 5 & 12 & 9 & 0.4 \\
        \hline
		2004 & 13 & 39 & 47 & 0.65 \\
        \hline
		2005 & 7 & 22 & 26 & 0.81 \\
        \hline
		2006 & 11 & 33 & 39 & 0.73 \\
        \hline
		2007 & 4 & 15 & 24 & 1.0 \\
        \hline
		2008 & 7 & 15 & 9 & 0.2 \\
        \hline
		2009 & 10 & 34 & 60 & 0.68 \\
        \hline
		2010 & 8 & 21 & 19 & 0.61 \\
        \hline
		2011 & 8 & 32 & 59 & 0.80 \\
        \hline
		2012 & 6 & 26 & 56 & 0.92 \\
        \hline
		2013 & 5 & 18 & 31 & 0.86 \\
        \hline
		2014 & 6 & 16 & 13 & 0.85 \\
        \hline
		2015 & 9 & 24 & 23 & 0.58 \\
        \hline
		2016 & 17 & 60 & 114 & 0.75 \\
		\hline 
	\end{tabular} 
    
	\caption{Overview of all NKOS papers sorted by years. In general, community shows a high average clustering in many years indicating that there are many triangles in the network.}
	\label{tab:workshops} 
\end{table}

%\fariba{please double check the table. Some numbers do not match the previous table. For example  now they are less nr. papers and authors in 2003 compared to previously reported table. Also for 2004 or 2007. Also please check the venue} 
%philipp{done! The reason was the restriction to co-authored paper! So we are fine here} 

%\begin{table}
%	\centering
%	\caption{Overview of all NKOS special issue papers. \fariba{do we still need this?}}
%\begin{tabular}{|c|c|c|}  
%	\hline 
%	venue& papers  & authors  \\ 
%	\hline 
%	JODI 2001 \cite{Hill2001} & 5 & 8 \\ 
%	\hline 
%	JODI 2004 \cite{Tudhope2004} & 5 & 15 \\ 
%	\hline 
%	NREV 2006 \cite{Tudhope2006} & 6 & 11 \\ 
%	\hline 
%	IJDL 2016 \cite{Mayr2016} & 7 &  20\\ 
%	\hline 
%\end{tabular} 
%\label{tab:SI}
%\end{table} 

%Table \ref{tab:SI} provides an overview of all papers in the special issues. We can see that a relative constant number of papers in the issues have an increasing number of authors. In average 2.3 authors published a special issue journal paper.

\section{Network analysis of the NKOS community}\label{analysis}

In order to analyze the collaboration of the NKOS community, we build a network of all authors at the workshops and special issues and compute various centrality measures for each author. A link in this network represents two authors who wrote a paper together.  Therefore, if we have $n_p$ number of papers and a paper \textit{i} has $m_i$ authors, the total number of pairs (links) $E$ are 

\begin{equation}
E = \sum\limits_{i=1}^{n_p} \frac{m_i(m_i-1)}{2} \qquad \qquad \qquad if\quad m_i\ge 1\\
\end{equation}

If two authors have published more than one paper together, we give weights to the link equivalent to the number of times they have collaborated in different papers. Thus, the resulting network is a weighted undirected graph.

In this paper, first, we investigate global properties of the network over time. Second, we analyze the centrality of the authors in this network. Lastly, we investigate gender differences in collaboration behavior in this community.

\section{Results}\label{results}

\begin{figure}
	\centering
	\includegraphics[width=0.8\linewidth]{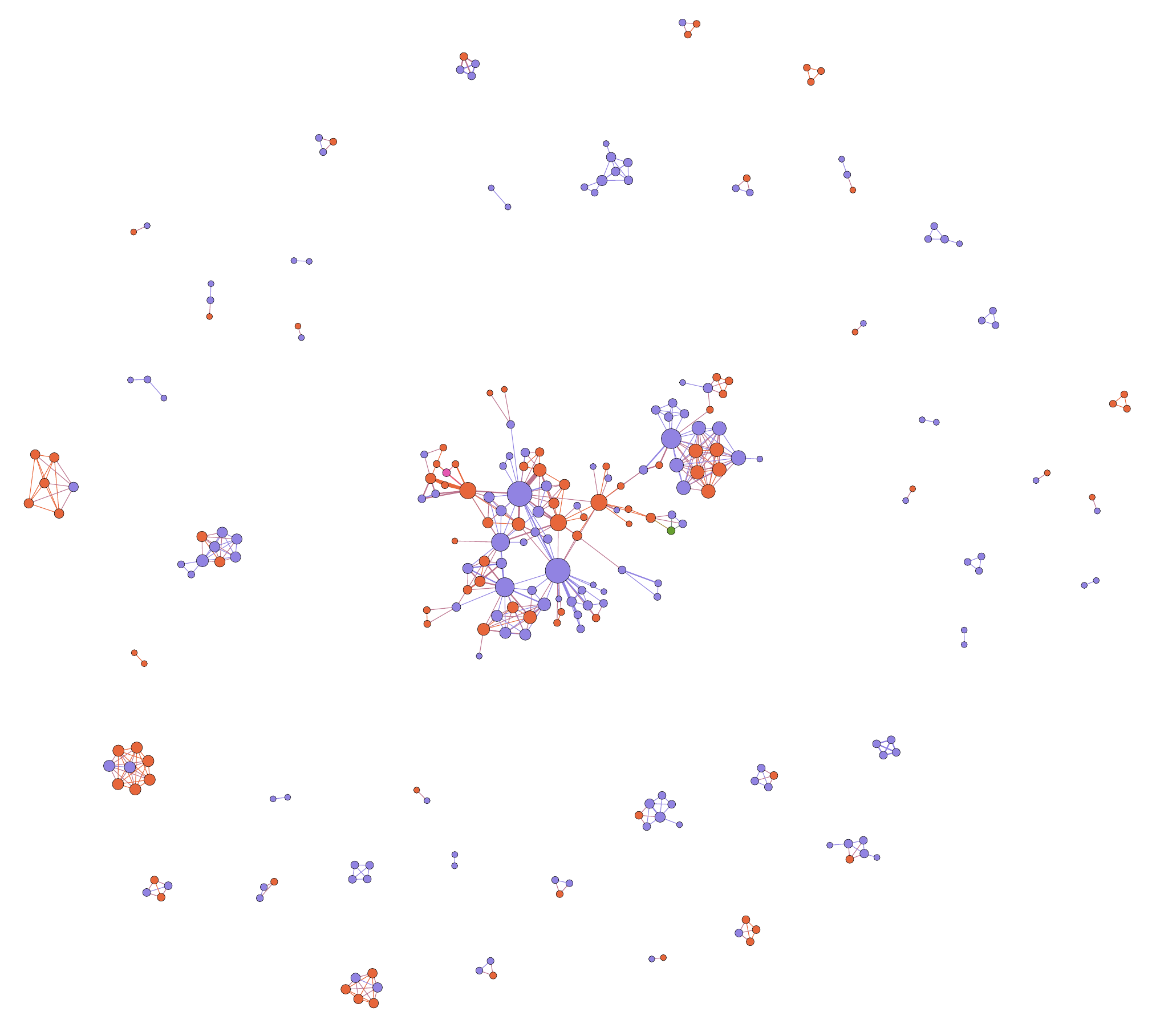}
	\vspace{-0.5em}
	\caption{Co-authorship network of the NKOS community. In general, the network is sparse and contains 44 isolated components. The largest connected component (the cluster in the middle) contains 107 number of nodes. Nodes are colored based on their gender. Purple nodes are men and orange nodes are women.}
	\label{fig:wholenet}
	\vspace{-0.5em}
\end{figure}

\begin{figure}
	\centering
	\includegraphics[width=0.9\linewidth]{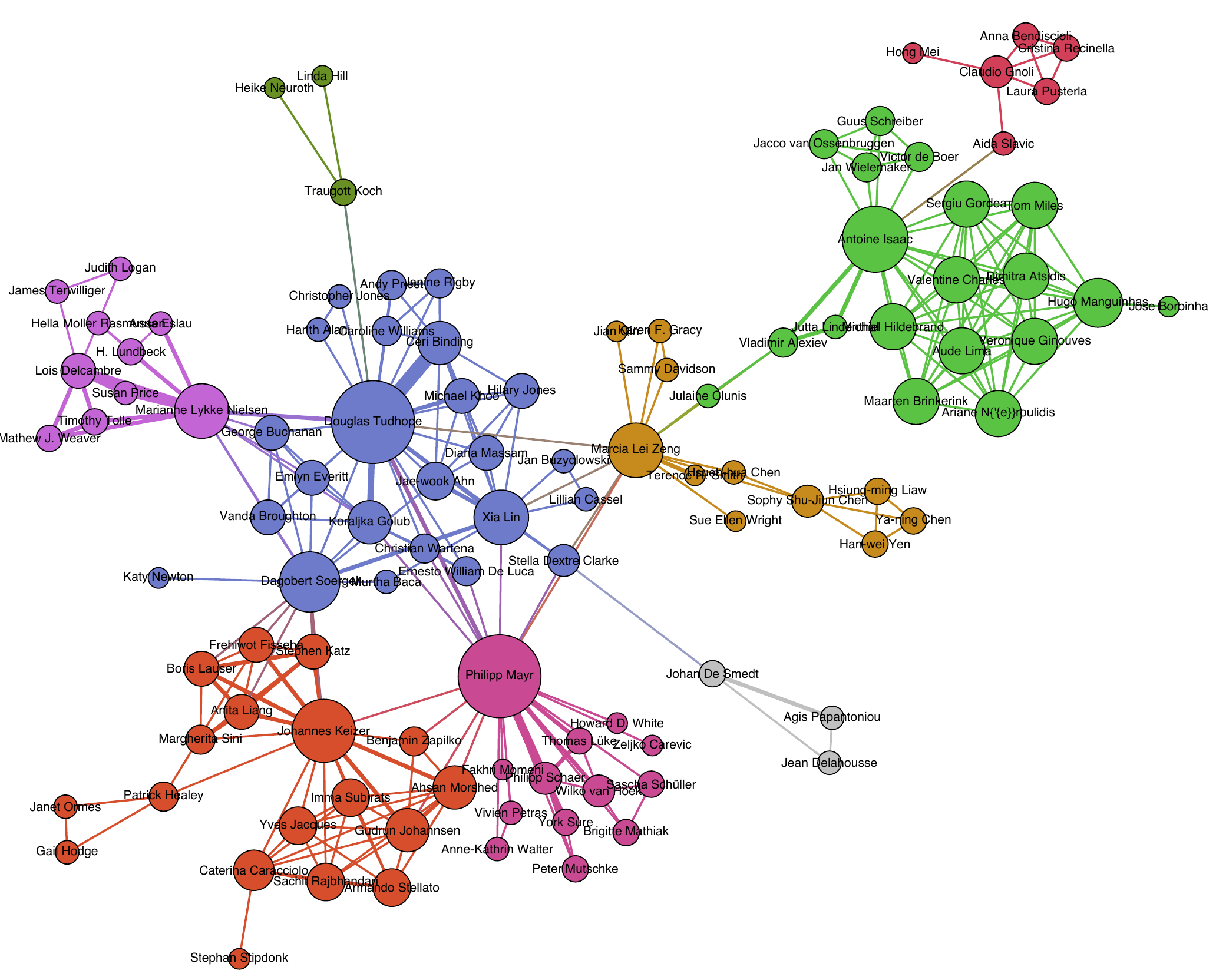}
	\caption{Largest component in the NKOS co-authorship network. The network is clustered into 9 clusters using Louvain clustering method \cite{blondel2008fast}. Nodes are colored based on their cluster and size of the node represents node's degree. Clusters are shaped based on the location of the groups and collaboration among members. Majority of the scholars in the largest component are based in Europe.}%\fariba{can we say something with regards to the formation of clusters? are they geographical? where is the US cluster?} %is it possible to produce a sequence of this graph, say the first 5 years (2000-2005), 2006-2010 and 2011-2016?
	\label{fig:largestComponent}
\end{figure}

Figure \ref{fig:wholenet} demonstrates the overall NKOS co-authorship network. In this view, each author has at least one co-author. The node color represents the gender; purple for men and orange for women. This network contains 44 components. From the network illustrated in this figure, we selected the largest component that is represented in Figure \ref{fig:largestComponent}. 107 authors (41\% of all authors) are connected in this component.  The NKOS co-authorship network in the "NKOS bibliography" is a typical co-authorship network with one relatively large component, some smaller components and many isolated co-authorships or triples.

\begin{figure}
	\centering
	\includegraphics[width=0.7\linewidth]{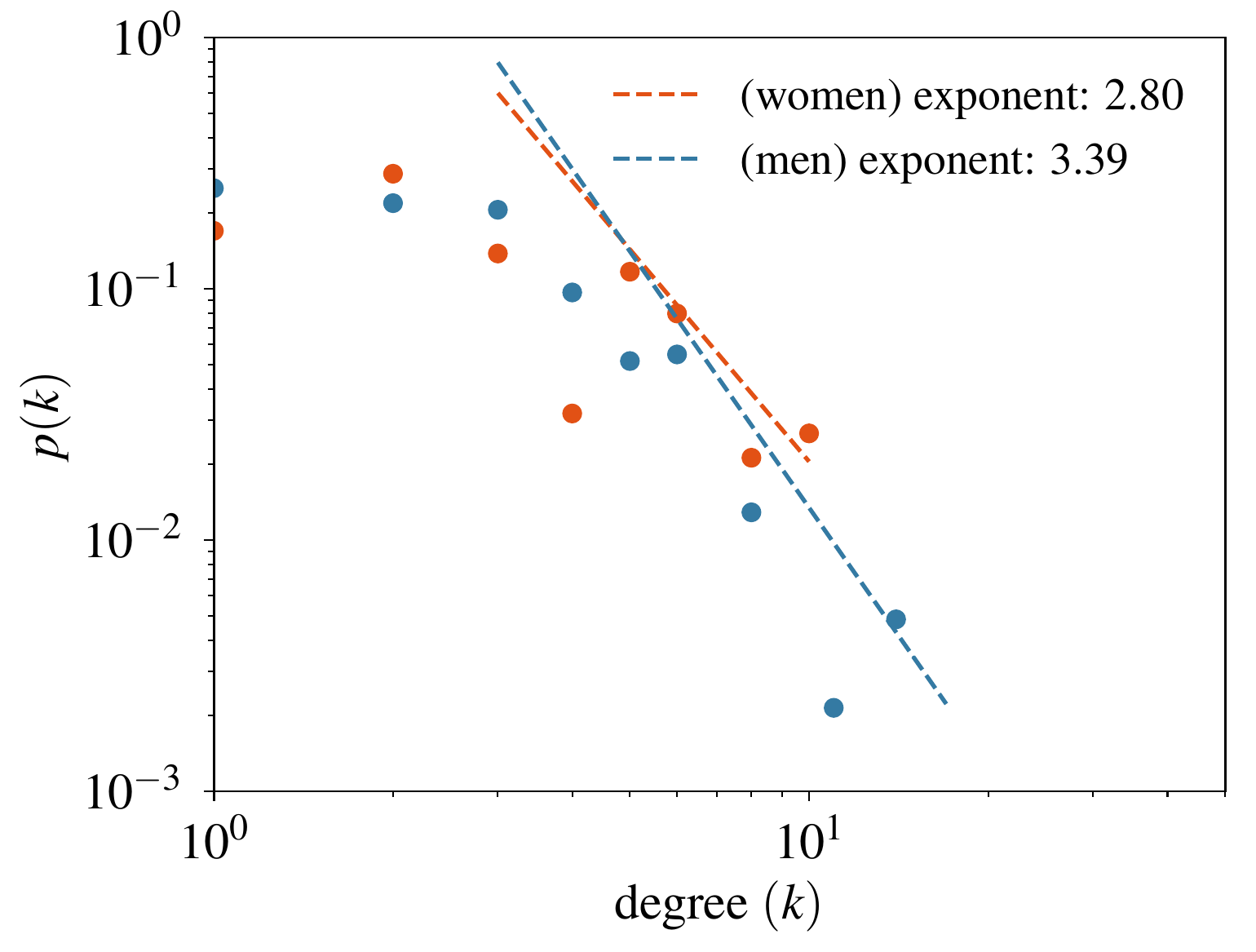}
	\caption{Degree distribution of the NKOS network. Blue and orange colors indicate the distribution for men and women respectively. Although the network is small, it exhibits power-law degree distribution.}
	\label{fig:Degreedistribution}
\end{figure}

Figure \ref{fig:Degreedistribution} shows the degree distribution for this network. Despite being a rather small network, the degree distribution follows a similar trend as a power-law degree distribution that has been observed in other co-authorship networks \cite{barabasi2009scale,jadidi2017gender}. 

In Figure \ref{fig:largestComponent}, the largest connected component, we see that scientists tend to forge intra-institutional collaborations \cite{evans_community_2011}. Good examples are the clusters from Johannes Keizer (FAO), Antoine Isaac (Vrije Universiteit Amsterdam/Europeana) and Philipp Mayr (GESIS). A large fraction of their co-authors are affiliated with the same institution. Also a tendency to select those co-authors who are in geographic proximity is visible in figure \ref{fig:largestComponent}. E.g. Douglas Tudhope (University of South Wales, UK) has a larger fraction of UK-affiliated co-authors.

\subsection{Node centralities}

%Some of the most commonly used centrality measures of nodes in a graph are degree, betweenness and closeness centrality. Degree is the number of nodes that a focal node is connected to and measures the total involvement of the node in the network \cite{Opsahl2010}. In our coauthorship network it indicates total number of unique collaborators that one author has. Betweenness centrality assesses the degree to which a node lies on the shortest path between two other nodes and is able to funnel the flow in the network \cite{Opsahl2010}. In this network the author with a high betweenness has a large influence in transfering the information from one part of the network to another. 

% To show the quantity of collaboration in the community we measured the degree centrality for each author. Figure \ref{fig:degreePercentage} shows the percentage of authors with different degrees. In this figure we see that 15\% (with degree=0) of authors had no co-authorship with others and 53\% of them had a maximum of 3 cooperations with other authors. 32\% had at least 4 co-authors for all their papers.

%\begin{figure}
%	\centering
%	\includegraphics[width=0.6\linewidth]{degreePercentage}
%	\caption{Distribution of degree numbers of authors in the network} 
%	\label{fig:degreePercentage}
%\end{figure}

\begin{figure}

\begin{minipage}{.5\linewidth}
\centering
\subfloat[]{\label{main:a }\includegraphics[scale=.5]{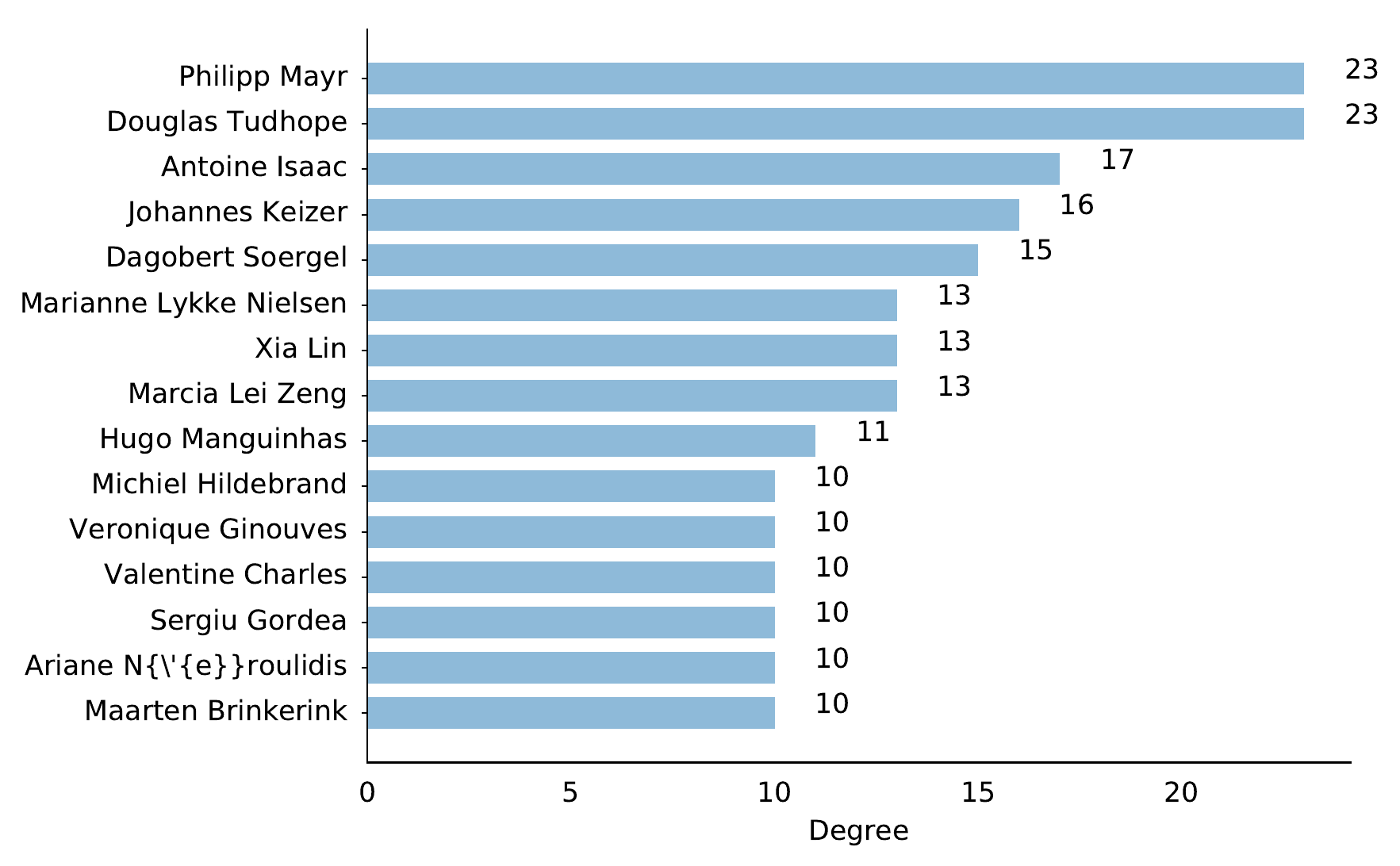}}
\end{minipage}\par\medskip
\begin{minipage}{.5\linewidth}
\centering
\subfloat[]{\label{main:b}\includegraphics[scale=.5]{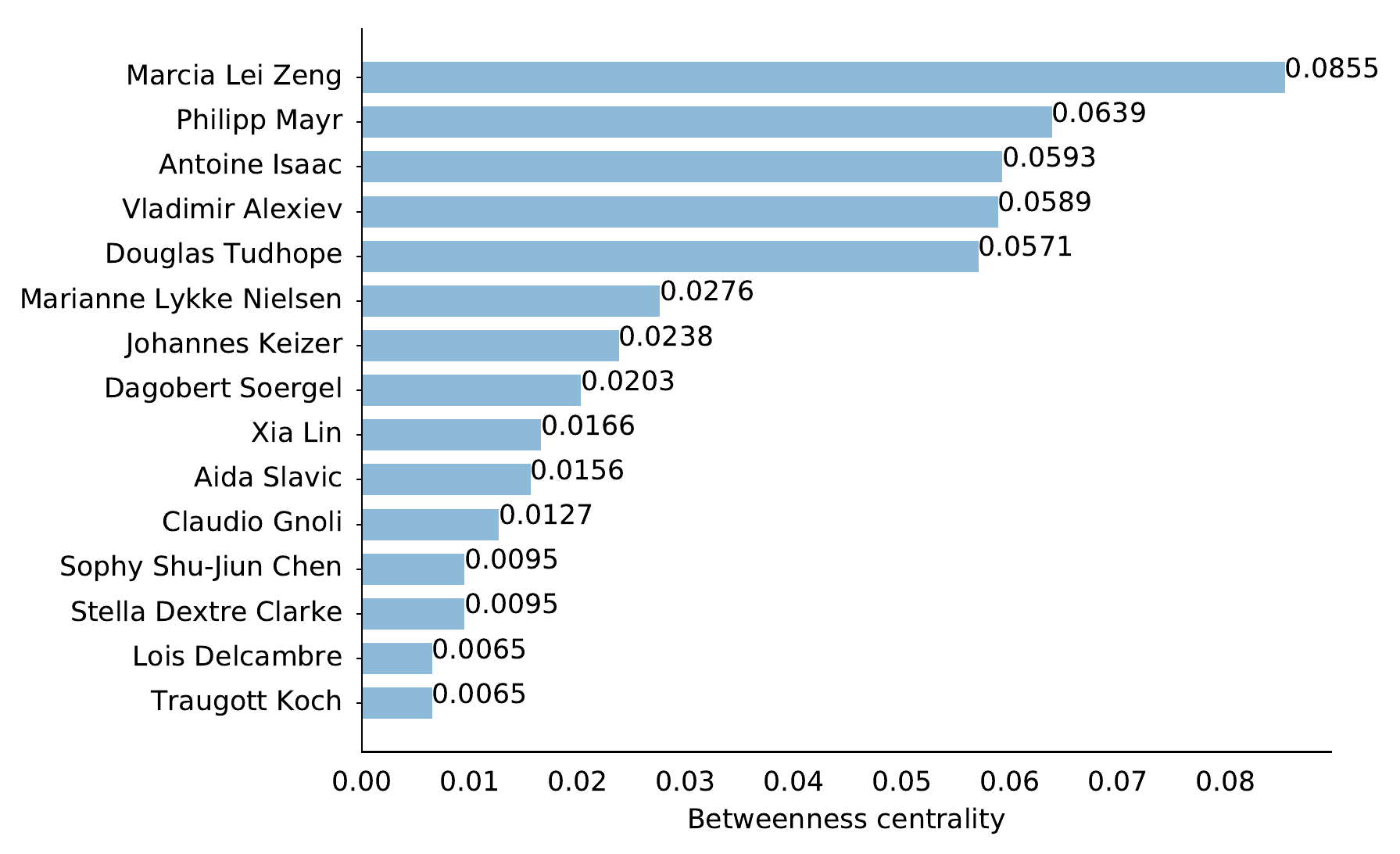}}
\end{minipage}\par\medskip
\begin{minipage}{.5\linewidth}
\centering
\subfloat[]{\label{main:c}\includegraphics[scale=.5]{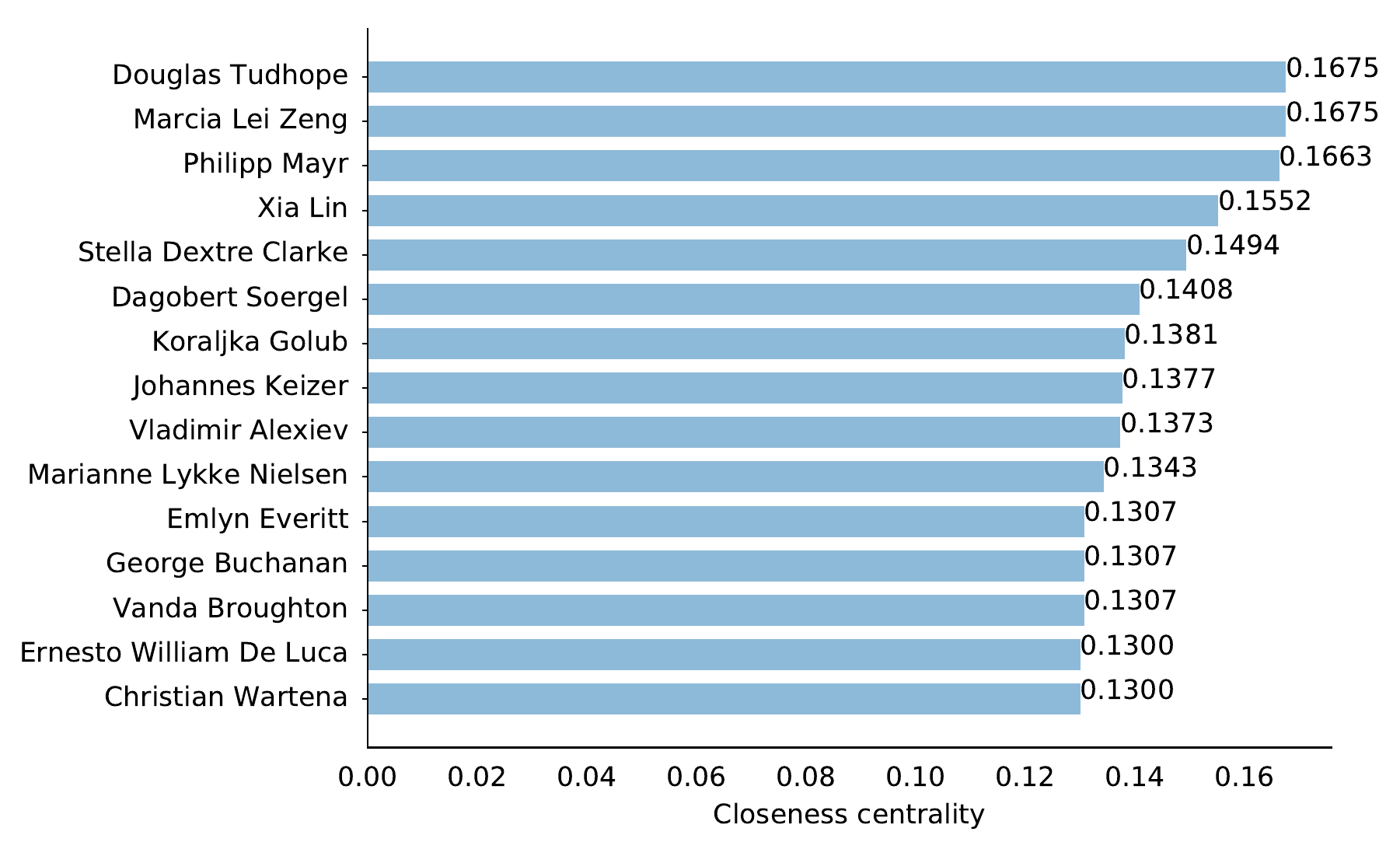}}
\end{minipage}

\caption{Top 15 authors with the highest (a) Degree centrality, (b) betweenness centrality and (c) closeness centrality.}
\label{fig:all_centrality}

\end{figure}

% \begin{figure}
% 	\centering
% 	\includegraphics[width=0.8\linewidth]{degree}
% 	\caption{Top 15 authors with highest degree centrality.}
% 	\label{fig:degree}
% \end{figure}

% \begin{figure} 
% 	\centering
% 	\includegraphics[width=0.8\linewidth]{betweenness.pdf}
% 	\caption{Top 15 authors with highest betweenness centrality.}
% 	\label{fig:betweenness}
% \end{figure}

% \begin{figure} 
% 	\centering
% 	\includegraphics[width=0.8\linewidth]{closeness.pdf}
% 	\caption{Top 15 authors with highest closeness centrality.}
% 	\label{fig:closeness}
% \end{figure}

To detect the influence of authors on information exchange we calculate various measures of centrality namely degree centrality, betweenness centrality and closeness centrality of the authors. Here we only focus on the largest connected component (LCC) in order to have a robust comparison.

Degree centrality is the most straightforward measure of centrality that depicts the importance of nodes in terms of total number of unique links. The authors with high degree centrality have established a wide collaboration with many different scholars. 

Betweenness centrality indicates fraction of shortest paths between all pairs of nodes that pass through a node. The betweenness of a node indicates the node's ability to funnel the flow in the network \cite{Opsahl2010}. In this network, the author with a high betweenness has a large influence in transferring the information from one part of the network to another. 

Closeness centrality indicates how close scholars are from others. Mathematically, it is sum of all the shortest paths between a node to all other nodes \cite{freeman1978centrality}. If a shortest path between node $u$ to $v$ is $d(u,v)$ and the total number of nodes in the graph is denoted by $N$, closeness centrality of the node $u$ is defined as follows:

\begin{equation}
c(u) = \frac{N-1}{\sum_{v}^{N-1} d(u,v)}
\end{equation}

where $N-1$ in the nominator normalizes the measure so that it becomes size independent. Scholars with high closeness centrality are on average closer to other nodes in the network.

%check http://journals.plos.org/plosone/article/file?id=10.1371/journal.pone.0059613&type=printable

Figure \ref{fig:all_centrality} shows the comparison of centrality measures for top 15 authors in the largest connected component. It is interesting to note that authors centrality ranks may vary depending on the type of the centrality measures. For example, even though H. Manguinhas has relatively high degree centrality, this author does not appear in the top closeness or betweenness rank. A closer look at the author's location in the graph \ref{fig:largestComponent} shows that this author is embedded in the light green cluster with high clustering and few connectivity with other clusters.

Comparing closeness centrality and betweenness centrality also shows interesting results. Although some authors have a high closeness to other scholars, they may not have high betweenness centrality. For example, K. Golub has a relatively high closeness centrality due to special location of the author in connection with many other authors from different clusters. However, this author does not have a relatively high betweenness centrality because her network position does not allow to connect other further distanced clusters. In contrast, author A. Slavic does not have a high degree or a high closeness centrality, but this author has a high betweenness centrality due to connecting an almost isolated red cluster to the rest of the network. The same is true for T. Koch. It is important to note that while scholars with higher closeness centrality are on average closer to other scholars and thus can access novel ideas more frequently, authors with high betweenness centrality play a crucial role in transferring the knowledge in the community \cite{iyer2013attack}. 

\subsection{Structural holes and bridges}
Weak ties play a crucial role in networks by connecting disconnected clusters and act as bridges in networks. Structural hole idea first coined by sociologist Ronald Burt, suggests that nodes can act as a mediator between two or more closely connected clusters. This is in particular important since novel ideas or information need to pass from these gatekeepers to transfer to other parts of the network. Here, we measure the effective size of a node based on the concept of redundancy. A person\textsc{\char13}s ego network has redundancy to the extent to which her neighbors are connected to each other as well. In a simple graph, the effective size of a node $u$, $e(u)$, can be expressed as:
\begin{equation}
e(u) = n - \frac{2t}{n}
\end{equation}
Where $t$ is the number of the total ties in the egocentric network (excluding those ties to the ego) and $n$ is the number of total nodes in the egocentric network (excluding the ego). The effective size can vary from 1 to the total number of links in the ego \cite{burt2004structural}. The higher the effective size, the more effective a node is in terms of being a bridge.

\begin{figure}
	\centering
	\includegraphics[width=0.8\linewidth]{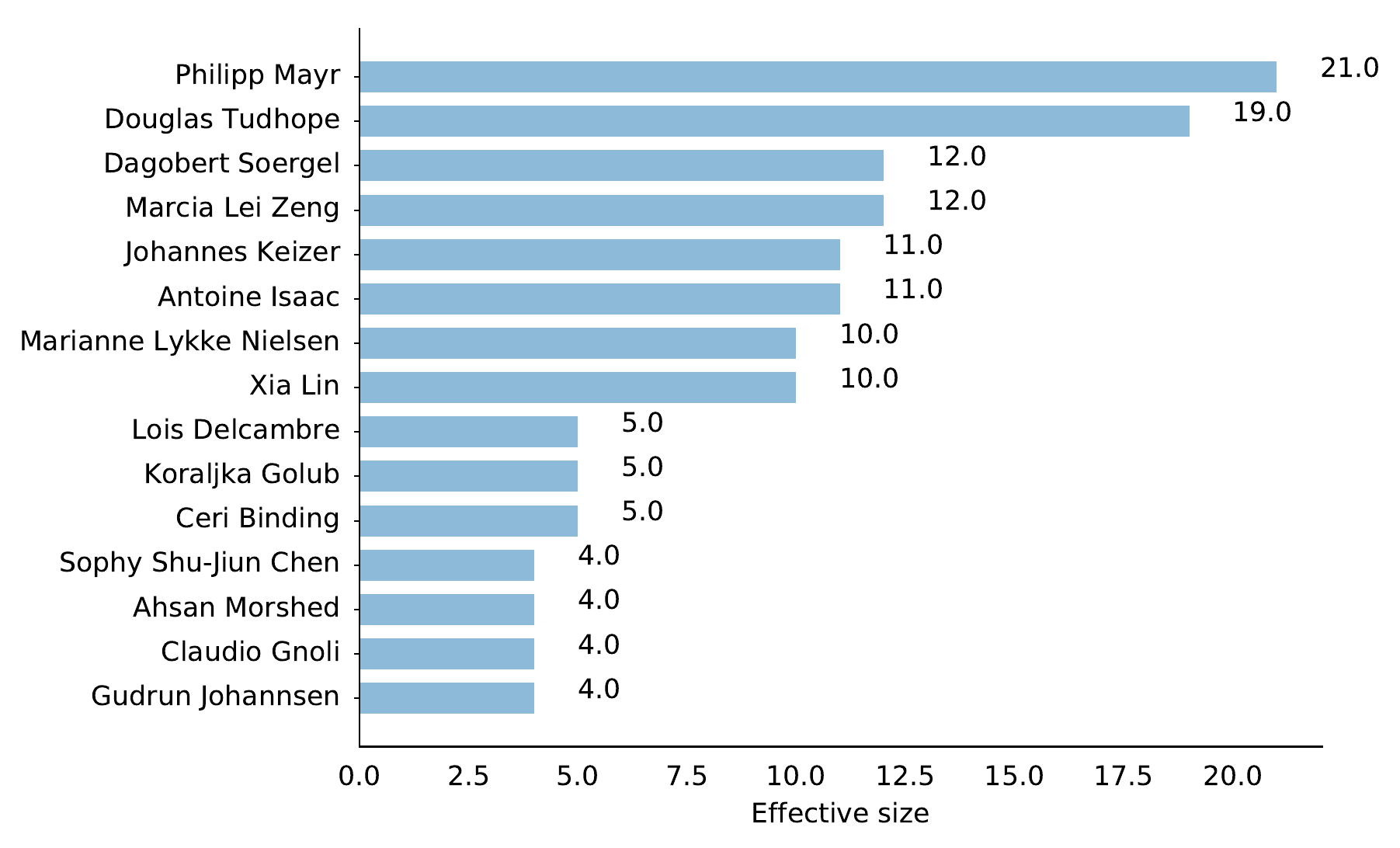}
	\caption{Top 15 scholars with the highest effective size. The effective size indicates the ability of a node to connect otherwise disconnected nodes and therefore the node can act as a weak tie or bridge.} 
	\label{fig:effective_size}
\end{figure}

Figure \ref{fig:effective_size} displays the top 15 ranked authors with respect to their effective size. The ranking suggests that in this community, nodes with high degree (hubs) also act as bridges between the clusters, thus, they can transfer novel ideas among their peers. 

\subsection{Gender differences in the co-authorship network}

To infer the gender of the scholars, we use the state-of-the-art approach by combining the results of the first names and Google images of the scholars with their full names \cite{karimi2016inferring}. For the remaining unidentified names or names with initials, we manually check the author's online profile based on the title of their papers. Our complete network consists of 97 (38\%) women and 157 (62\%) number of men and 2 unidentified names. Compared to other scientific communities and in particular in science and engineering fields, this community shows a higher percentage of active women \cite{jadidi2017gender}. The share of women and men in the largest connected component also shows an interesting effect. We find 46 women and 59 men in the LCC which means women occupy 43\% of the nodes in this component. 

\paragraph{Homophily.}
In the first step, we measure homophily in this network. There are various ways to define homophily. Here, we use two well-defined measures. 
First measure of homophily is proposed by Newman that computes the Pearson correlation between attributes when corrected by what we would expect from node's degree \cite{newman2002assortative}. The homophily varies between -1 (disassortativity) to +1 (complete assortativity). We find that gender assortativity in this community is 0.1. This means that there is a positive tendency among scholars in this community to collaborate with similar gender. One can observe the gender homophily from figure \ref{fig:wholenet}.

\begin{figure}
	\centering
	\includegraphics[width=0.99\linewidth]{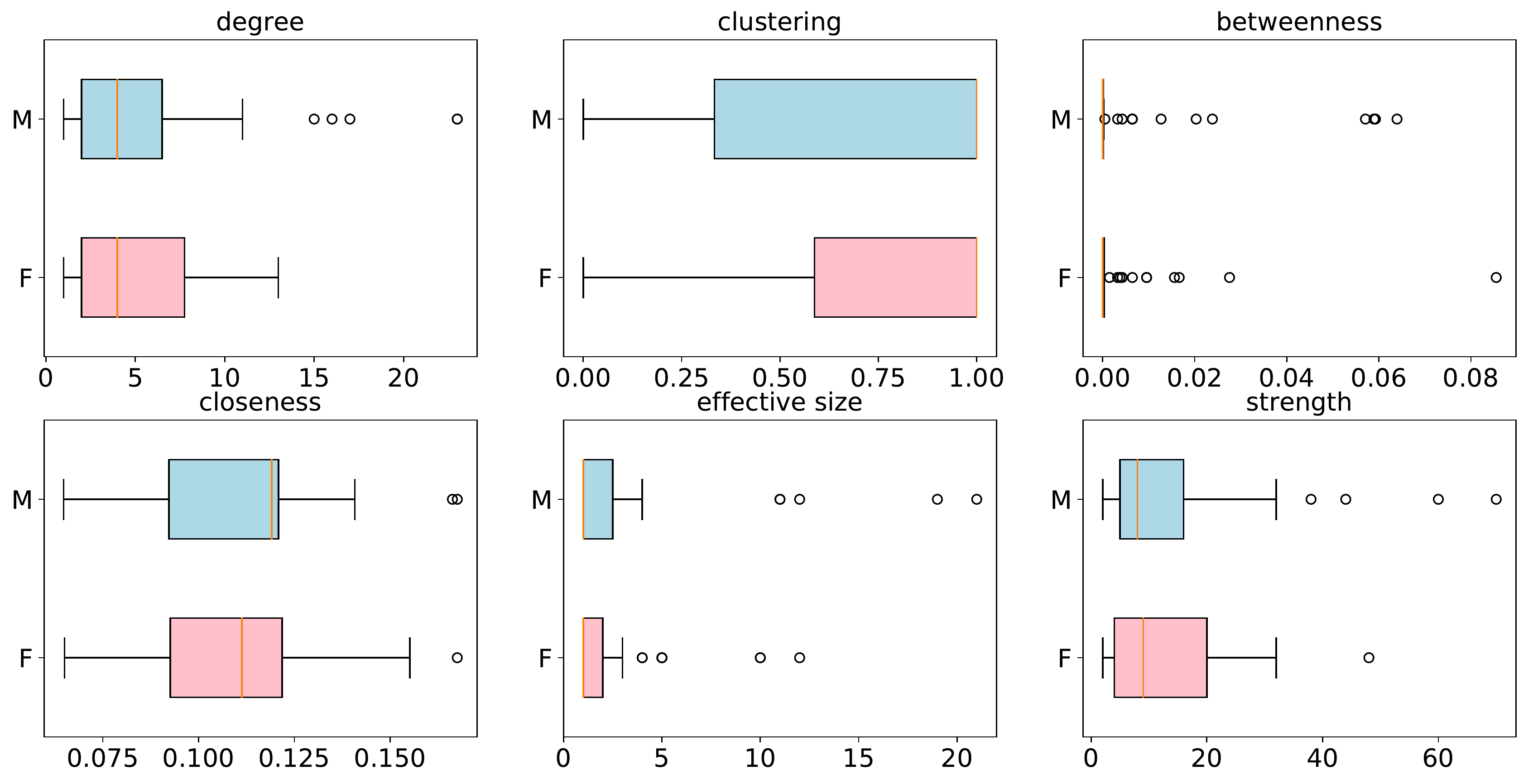}
	\caption{Box plots indicating median and quartiles of network properties for male and female scholars in the largest connected component. Median is similar for majority of the node characteristics except for closeness centrality that is higher for men. With regards to degree centrality there are more outliers among men with high degree. For clustering, women have higher clustering on average than men. Men also show outliers with higher effective size and strength compared to women.} 
	\label{fig:gender_effect}
\end{figure}

Although the assortativity measure captures the overall homophily in the network, it does not provide additional insights whether or not the nature of homophily is symmetric or asymmetric. Indeed, we have shown previously that asymmetric homophily can impact the degree centrality of the nodes and in particular minority group in networks \cite{karimi2017visibility}. To capture the asymmetric nature of the homophily, we take a simple approach first proposed by Coleman (1958). In this case we measure the probability of links that exist between two scholars of the same gender. Let us denote the probability of links that exist among women as $p_{ww}$ and among men as $p_{mm}$. To compare groups of different sizes, the probabilities are compared with group sizes and normalized by the maximum values. If the fraction of women is denoted by $f_{w}$ and men by $f_{m}$, the Coleman index for women is:

\begin{equation}
C_{w} = \frac{p_{ww}-f_{w}}{1-f_{w}}
\end{equation}

Similar definition will apply for men. The maximum value for Coleman homophily index is 1. When applying this index to our network we get $C_{w} = -0.12$ for women and $C_{m} = -0.42$ for men. These results suggest that the homophily among women is higher than the homophily among men in this network. Similar findings were also found in other co-authorship networks \cite{jadidi2017gender}. %It worths mentioning that the asymmetric homophily can have non-trivial consequences on the centrality of the groups and the degree distribution \cite{karimi2017visibility}. 

\paragraph{Network characteristics and gender differences.} Next, we measure the network characteristics among men and women in the largest connected component. We use six measures of networks similar to the previous section. We also include strength of the node as the sum of all wighted links. 

Figure \ref{fig:gender_effect} shows box plots comparing network measures for men and women. Overall, the median and quartiles for degree and betweenness are the same for men and women. Women show higher tendency for higher clustering compared to men. Men show higher median for closeness centrality compared to women. In addition, there is a higher number of outliers among men in terms of the degree, effective size and strength compared to women.

%todo add some more details and interpretation

\section{Conclusion}\label{concl}
In this paper, we have analyzed the collaborative research of authors and their connectivity for the special case of NKOS workshop activities including four special issues on NKOS. The results highlight the most active and central scholars in this community. We found differences among centrality measures of the scholars which indicate that scholars play a different role in their collaboration network. We also found the most influential scholars who act as bridges between the clusters. We found 9 clusters in the largest component that show scholars have higher tendency to collaborate with those in the same institution or the same geographic proximity \cite{evans_community_2011}. Our analyses show that NKOS community is rather successful in bringing researchers from different domains together in recent years.

NKOS co-authorship network consists of 38\% women in total, and the share of women in the largest connected component is 43\%. The network shows positive gender homophily and the homophily among women is higher compared to men. We found on average men have higher closeness centrality compared to women. In addition, women have slightly higher clustering compared to men. Apart from these differences, we do not find any significant dissimilarities between men and women with respect to their centralities.

%We saw some details of the largest connected component in this network that covers 41\% of the authors of the whole network. 

%The analysis of the NKOS community shows some typical tendencies of researcher networks and their collaboration pattern \cite{evans_community_2011}. 

%We can see that NKOS researchers tend to work in intra-institutional collaborations. Also a tendency to select co-authors who are in geographic proximity is visible in the largest connected component (see Figure \ref{fig:largestComponent}) and smaller components (see Figure \ref{fig:wholenet}).

%todo: add some interpretation

This study has some limitations. First of all, we have included just research paper presentations. Editing and organizing activities at the workshops, which have an enormous impact on the visibility and connectivity of researchers, have not been covered in our dataset. This leads to artifacts, e.g. Traugott Koch,\footnote{Traugott Koch was an central protagonist and networker of the U.S. and European NKOS community. He retired and left the NKOS community in 2012.} a long-term organizer of the NKOS workshops and editor of the early JoDI special issues on NKOS, is not covered very well in our dataset and the network. 

Second, many influential papers (e.g. \cite{hodge2000,Zeng2004}) and standardization activities (e.g. the W3C Recommendation for SKOS \cite{miles_skos_2009}), presented and discussed at NKOS events and published after the NKOS workshops are missing. This fact is reducing the representativeness and completeness of the network.

Third, we have not included bibliometric data to complete our analysis. This is because most of the NKOS workshop activities (presentations) are not formally cited or even mentioned in scientific papers. In difference to the workshop output, the few journal papers in the special issues on NKOS are cited. Some works (e.g. \cite{SPCranefield2001,SPDoerr2001,SPTudhope2001,SPSoergel2004,SPTrant2006}) are cited well in the literature. So adding citation data would be a next reasonable step to complete the dataset.

\section{Future work}\label{future}
We are planning to extend the analysis of the NKOS network. In this way, we first plan to complement the dataset with other NKOS research output. We also plan to analyze the development of topics in the titles and abstracts of the presentations and papers. Combining network analytic measures with bibliometric analysis (e.g. co-citations, bibliographic coupling) would complement our preliminary observations and advance our understanding of the role of gender and other attributes in scientific collaboration. We invite people to contribute to our open dataset.

\section{Acknowledgment}\label{sec:ACKNOWLEDGMENTS}

We thank our colleague Marcia Lei Zeng (Kent State University) who provided us with internal information about the U.S. NKOS workshops. 
This work was partly funded by DFG, grant no. SU 647/19-1; the "Opening Scholarly Communication in the Social Sciences" (OSCOSS) project at GESIS. 

%\newpage

\bibliographystyle{splncs03} % abbrev
\bibliography{ijdl2018} 

\end{document}